\RequirePackage{lineno}
\documentclass[superscriptaddress,amsmath,amssymb,nofootinbib]{revtex4}

\usepackage{graphicx}% Include figure files
\usepackage{dcolumn}% Align table columns on decimal point
\usepackage{bm}% bold math
\usepackage{ulem}
\usepackage{xfrac}

\usepackage[printwatermark]{xwatermark}
\usepackage{xcolor}

\usepackage{graphicx,amsmath,amssymb}
\usepackage{epstopdf}
\usepackage{lineno}
\usepackage[colorlinks=true, urlcolor=blue, linkcolor=blue, citecolor=blue]{hyperref}

\begin{document}

\title{Effect of gravitational field self-interaction on large structure formation}

\author{ Alexandre Deur}

\affiliation{University of Virginia, Charlottesville, VA 22904. USA\\
\email{deurpam@jlab.org} 
}


\begin{abstract}
We check whether General Relativity's field self-interaction alleviates the need for dark matter to explain the 
universe's large structure formation. We found that self-interaction accelerates sufficiently the growth of structures so that
they can reach their presently observed density. No free parameters, dark components or modifications of the known laws
of nature were required. 
This result adds to the other natural explanations provided by the same approach to the, {\it inter alia}, flat rotation 
curves of galaxies, supernovae observations suggestive of dark energy, and dynamics of galaxy clusters,
thereby reinforcing its credibility as an alternative to the dark universe model. 
\end{abstract}
%\keywords{General relativity, QCD, Dark Matter, Dark Energy}

\maketitle

\section{Introduction}

The essential role of dark matter in the growth of galaxies and other large structures %of the universe 
constitutes an important 
evidence for the reality of that dark component of the universe. Cosmic microwave background  (CMB) anisotropy data show~\cite{CMB}
that at recombination time ($t \approx 3.7\times10^{-4}$ Gyr, or redshift $z \approx 1100$), 
the fractional density  
fluctuations $\delta$ that will evolve to form the large structures have typical  magnitudes of $ \approx10^{-5}$. Modeling their evolution 
{\it via} the Jeans collapse mechanism  and assuming solely baryonic matter yield, for the present times, $\delta \approx 10^{-2}$. This is
2 orders of magnitude lower compared to observations. Dark matter solves this problem since its lack of
electromagnetic interaction allows it to start to coalesce without impediment from electromagnetic pressure, and therefore 
significantly earlier than visible matter whose growth is then accelerated by the relatively denser dark matter halos. 

Although this model is generally successful in describing the distribution of the universe's matter density, it predicts
too many dwarf galaxies and globular clusters~\cite{Klypin:1999uc}. 
Furthermore, unease is growing from the absence of direct~\cite{Kahlhoefer:2017dnp} or 
indirect~\cite{Gaskins:2016cha} detection of dark matter particles, especially since these searches have largely 
exhausted the parameter space of
the most natural and compelling theories, e.g. SUSY, that provide suitable candidates~\cite{Arcadi:2017kky}. 
These problems motivate theories alternate to dark matter. While in principle those need not solve all 
the questions involving dark matter single-handedly, solving these questions consistently would strongly support the plausibility of an alternate theory.
Foremost among the questions to be addressed is the growth of large structures. 
In this article, we discuss whether the self-interaction (SI) of gravitational fields, a well-known feature of General Relativity (GR),
may help resolve the problem of the too slow growth of large structures,
in the absence of dark matter.
Without modifying  the known laws of nature or assuming exotic matter, GR's SI already provides natural solutions for several 
observations linked to dark matter, namely 
(1) the flat rotation curves of galaxies~\cite{Deur:2009ya}-\cite{Deur:2020wlg}; 
(2) the tight empirical relation between the accelerations measured in galaxies and the accelerations calculated from the galactic baryonic contents~\cite{Deur:2019kqi};
(3) the internal dynamics of galaxy clusters, including the Bullet Cluster~\cite{Deur:2009ya};  
(4) the Tully-Fisher relation~\cite{Deur:2009ya, Deur:2017aas}; and
(5) the correlation between the ellipticity of early-type galaxies and their inferred dark mass~\cite{Deur:2013baa}. 
Furthermore, and especially relevant for this article, when GR's SI are accounted for in the universe's evolution equation, no 
%cosmological constant or 
dark energy is needed~\cite{Deur:2017aas}. Because the GR's SI approach 
connects straightforwardly dark matter and dark energy  (see next section), 
the formalism used in Ref.~\cite{Deur:2017aas} can be directly applied to the problem of structure formation. 

In the next section, we recall the origin of GR's SI and summarize the similarities between  GR and the
Strong Interaction, whose SI effects are well-known and can thus guide the study of GR's SI. 
Next, using the global formalism of Ref.~\cite{Deur:2017aas}, we show how GR's SI accelerates the growth 
of large structures, thereby offering an explanation of their
formation without requiring dark matter. 
%We then discuss the closely related question of the structure distribution in space. 
%We then summarize our findings.

\section{Field self-interactions in General Relativity and the Strong Interaction }
Besides their fundamentally  different interpretation of the nature of gravitation, GR and Newton's gravity also
crucially differ in that GR is non-linear. This can be traced to field SI once space-time curvature is interpreted in terms of 
fields. GR can be formalized by the Einstein-Hilbert Lagrangian density, 
\begin{eqnarray}
\mathcal{L}_{\mathrm{GR}}={(\det~g_{\mu\nu})^{\sfrac{1}{2}}\, g_{\mu\nu}R^{\mu\nu}}/{(16\pi G)}, 
\label{eq:Einstein-Hilbert Lagrangian}
\end{eqnarray}
with
$g_{\mu\nu}$ the metric,
$R_{\mu\nu}$ the Ricci tensor and
$G$ the gravitational constant.
The gravitational field $\varphi_{\mu\nu}$ originating from a unit mass source is the variation of $g_{\mu\nu}$ 
with respect to a constant metric $\eta_{\mu\nu}$:
$\varphi_{\mu\nu}= (g_{\mu\nu} - \eta_{\mu\nu})/\sqrt{M} $, where $M$ is the system mass. 
Expanding $\mathcal{L}_{\mathrm{GR}}$ in term of $\varphi_{\mu\nu}$ yields, in the pure field case~\cite{Zee}:
\begin{eqnarray}
\mathcal{L}_{\mathrm{GR}}\!=\!\left[\partial\varphi\partial\varphi\right]\!+\!\sqrt{16\pi MG}\left[\varphi\partial\varphi\partial\varphi\right]\!+ 
\!16\pi MG\left[\varphi^{2}\partial\varphi\partial\varphi\right]\!+\cdots ,
\label{eq:Polynomial Einstein-Hilber Lagrangian}
\end{eqnarray}
where $\left[\varphi^{n}\partial\varphi\partial\varphi\right]$ represents a sum of Lorentz-invariant terms of the form $\varphi^{n}\partial\varphi\partial\varphi$.
Newton's gravity is given by $\mathcal{L}_{\mathrm{GR}}$ truncated to $n=0$, with 
$\left[\partial\varphi\partial\varphi\right]=\partial^{\mu}\varphi_{00}\partial_{\mu}\varphi^{00}$, $\eta_{\mu\nu}$ the flat
metric and $\partial^0\varphi_{00}=0$. The $n>0$ terms induce field  SI.
%%%ESTIMATE for importance of the non-linear effect:
\iffalse
The dimensionless coupling is $\sqrt(16\pi GM/l)$ with
$M$ the system mass and
$l$ a characteristic length of the system.
The non-linear effects grow proportionally to the radius of the galaxy-\cite{Deur:2020wlg}, therefore the dimensionless
parameter characterizing the non-linearity at the edge of a galaxy of radial size $R$ is:
$\alpha=\sqrt(16\pi GM(l')/l) R/l'$, where $l' \ll R$ is the length over which the field SI mostly occur. Since the galactic mass at small
radius is dominated by the core, we can take $l'$ to be core radius $l' \approx r_c$. Furthermore assuming $l=l'=r_c$, we have
$\alpha= \sqrt(16\pi GM(r_c)/r_c) R/r_c$.
Taking $M(r_c)=10^8~M_\odot$, $r_c=15$ pc and $R=3\times10^4$.
$\alpha \approx 8$ !!!TO BE CHECK!!!!
Assuming instead $l=r_0$, the characteristic length of the galaxy, typically 1 or 2 $kpc$, 
yields $\alpha \approx 1$. Taking {\it overly} conservatively $l=R$ still yields $\alpha \approx 0.2$.
In all case, the values of $\alpha$ shows that these effects are important.
%%NOTE THAT WE DON"T ACCOUNT FOR THE SUPPRESSING EFFECT OF THE SPHERCAL SYMMETRY%%
\fi

Another fundamental force displaying  SI is the Strong Interaction. It is formalized by quantum chromodynamics 
(QCD) whose pure field Lagrangian is:
\vspace{-0.3cm}
\begin{eqnarray}
\label{eq:QCD Lagrangian}\mathcal{L}_\mathrm{{QCD}}\!=\!\left[\partial \phi \partial \phi \right]\!+\!\sqrt{\pi \alpha_s}\left[\phi^2 \partial \phi \right]\!+\!
\pi \alpha_s\left[\phi^4\right],
\end{eqnarray}
Here, $\phi_\mu$ is the gluonic field and $\alpha_s$ the QCD coupling~\cite{Deur:2016tte}. 
Again, a bracket $[~]$ indicates a sum of Lorentz-invariant terms, and, in the QCD case,
contractions of the color indices. 
The similarity between $\mathcal{L}_{\mathrm{GR}}$ and $\mathcal{L}_{\mathrm{QCD}}$ makes the latter 
useful as a guide for GR in its strong regime, since QCD in that regime is well-studied. 
Like GR, the terms beyond $\left[\partial \phi \partial \phi \right]$ induce the field SI. 
They are interpreted in QCD (GR) as arising from the color charges (energy-momentum) 
carried by gluonic (gravitational) fields, which permit field self-coupling.
For QCD the coupling --driven by $\alpha_{s}$--
is large, making the consequences of field SI prominent.
In GR, large $\sqrt{GM/L}$ values ($L$ is a characteristic length of the system) enable SI. 
QCD's SI strongly increases the interaction between color charges and causes quark confinement. 
Likewise, GR's SI increases the gravitational system's binding 
compared to  Newton's theory. If the latter is used to analyse  
 galaxy or cluster dynamics, as is commonly done, ignoring the SI-induced intensification of the force 
 then creates a missing (dark) mass problem. 
The results of Refs.~\cite{Deur:2009ya}-\cite{Deur:2020wlg} %based on different non-perturbative approaches to compute GR'SI
indicate that SI can sufficiently  strengthen the gravitational binding such that no dark
matter is required to explain galactic rotation curves or the internal dynamics of  
galaxy clusters. %, including that of the Bullet cluster~\cite{bullet cluster}.
In QCD, the SI strengthens so much the binding of color sources that they remain confined, i.e. 
the Strong Interaction is essentially\footnote{The 
residual Strong Interaction  outside a nucleon, e.g. the Yukawa
interaction, is much weaker than the quark-quark interaction.} suppressed outside of the system, e.g. outside a nucleon.
This can be globally understood from energy conservation: the confined field increases the system's binding energy, but
the field concentration causes its depletion outside of the system. 
Likewise\footnote{Although energy conservation does not always hold in GR, it does for localized systems such as galaxies or
galaxy clusters.} in gravitational systems, the increased binding due to GR's SI weakens gravity's action at large scale.
If GR's SI is ignored, this weakening can then be misinterpreted as a large-scale repulsion, {\it viz} dark
energy.  The effect is time-dependent: as massive structures form, some gravitational fields become trapped in them, weakening
their manifestation at  larger scale. This implies a direct connection between dark energy and dark matter, particularly
between dark energy and the onset of structure formation.

In summary, GR and QCD possess similar Lagrangians whose structure enables field SI.
Those are prominent for QCD because $\alpha_s$ is large. Analogously, they must become important for GR 
once the system's $\sqrt{GM/L}$ is large enough. Performing a Newtonian analysis for massive systems, or using GR assuming
isotropy and homogeneity\footnote{While the universe's evolution equation derives from GR, the standard approximations of isotropy and homogeneity suppress the SI~\cite{Deur:2009ya, Deur:2019kqi, Deur:2013baa}. This is manifested by the ability  to model
with surprising accuracy the universe evolution using Newton's theory~\cite{Milne:1934}.} 
overlooks the effects of SI, resulting in an apparent missing mass inside the systems and an apparent 
global repulsion at larger scale. 
The apparent missing mass and global repulsion can then be interpreted as dark matter and dark energy, respectively. 
The GR SI approach is supported by parallels between QCD's phenomenology and the observations involving dark matter and dark energy, e.g. 
the similarity between the hadrons' Regge trajectories~\cite{the:Regge} and the galactic Tully-Fisher relation~\cite{Tully:1977fu}, or the
stronger (weaker) force manifestation inside (outside) systems, those being either hadrons, galaxies or galaxy clusters. The approach is discussed in detail and quantitatively 
in~\cite{Deur:2009ya}-\cite{Deur:2013baa}. Here, we study whether it can also explain 
large structures' growth.

\section{Role of field self-interaction in large structure formation}
As summarized above, GR's SI increases the internal binding of a massive system, i.e  the
interaction between components of the system. 
In the method described in Refs.~\cite{Deur:2009ya,Deur:2016bwq} (a lattice numerical approach), 
the force strengthening is interpreted as the collapse of field lines into the galactic disk, effectively reducing the 3-dimensional (3D)
force to a 2-dimensional (2D) force $F_{\mathrm{2D}}\propto1/r$.  
This interpretation mirrors that of QCD for which lattice calculations of static two-body systems show 
that field lines collapse into 1-dimensional (1D) structures called QCD strings
%\footnote{QCD strings were anticipated on phenomenological basis long 
%before the advent of lattice QCD %by Nambu, Nielsen and Susskind 
%to explain Regge trajectories~\cite{Nambu:1974zg}.}
 or flux-tubes~\cite{Bali:1994de}. The collapse induces a constant force
$F_{\mathrm{1D}}$ between the two bodies ({\it viz}, the static quarks) since the field lines cluster along the segment linking the two bodies. 
In the disk case, the $F_{\mathrm{2D}}\propto1/r$ force  results from the field lines collapsing in the approximately 2D axially
symmetric disk. %, which leads automatically to flat rotation curves~\cite{Deur:2009ya,Deur:2016bwq}. 
Finally, a system exhibiting a 3D spherical symmetry would retain a $F_{\mathrm{3D}}\propto1/r^2$ force 
since field lines have no preferred direction of collapse, {\it viz}
there would be no SI strengthening of the force. 
%This suggests a positive correlation between the ellipticities of elliptical galaxies  
% and their inferred dark masses, a prediction that was subsequently verified~\cite{Deur:2013baa}. 
In the QCD case, the 1D system displaying $F_{\mathrm{1D}}=\mbox{const.}$ 
is composed of a static quark and a static anti-quark. 
In the gravitational case, a 1D system massive enough to trigger GR's SI would be that of two galaxies.
In that case, interactions will be stronger than the Newtonian expectation\footnote{For this to occur, 
the 2-galaxy system should contain at least one elliptical galaxy. Otherwise, field 
line collapse would happen within the galaxies themselves, thereby trapping most of their 
gravitational field within their structures, leaving no field outside for the galaxy-galaxy interaction. 
%This would happen if the two galaxies are massive disk galaxies. 
Thus, we expect reduced interactions 
between two massive disk galaxies, and enhanced interactions otherwise.}
and string-like structures akin to QCD's flux tubes 
should arise at the intergalactic/galactic cluster scale.
This qualitatively agrees with the web structure of the present-time universe
and the necessity of enhanced gravitational interaction (or mass density) to explain the growth rate of large structures. 
In Ref.~\cite{Deur:2020wlg}, a background field method is used to compute GR's SI effects. %in disk galaxies. 
The background field represents the total gravitational field 
of the system and is treated as a space-time curvature. 
The individual interaction between two particular bodies composing the system 
is treated as a traditional force ({\it viz} a field line density). Just like for light lensing, the space-time curvature 
focuses the gravitational field lines between the individual bodies forming the system. This increases the field line density
i.e. increases  the force,
like light lensing magnifies the background object's luminosity. 
In the case of inhomogeneous systems made of multiple point-like bodies, lensing happens when three bodies are approximately 
aligned. The alignment is then strengthened and further enhanced, consistent with the previous discussion based on the lattice method and again suggestive of the universe's web structure. 

\begin{figure}
\center
\includegraphics[width=0.4\textwidth]{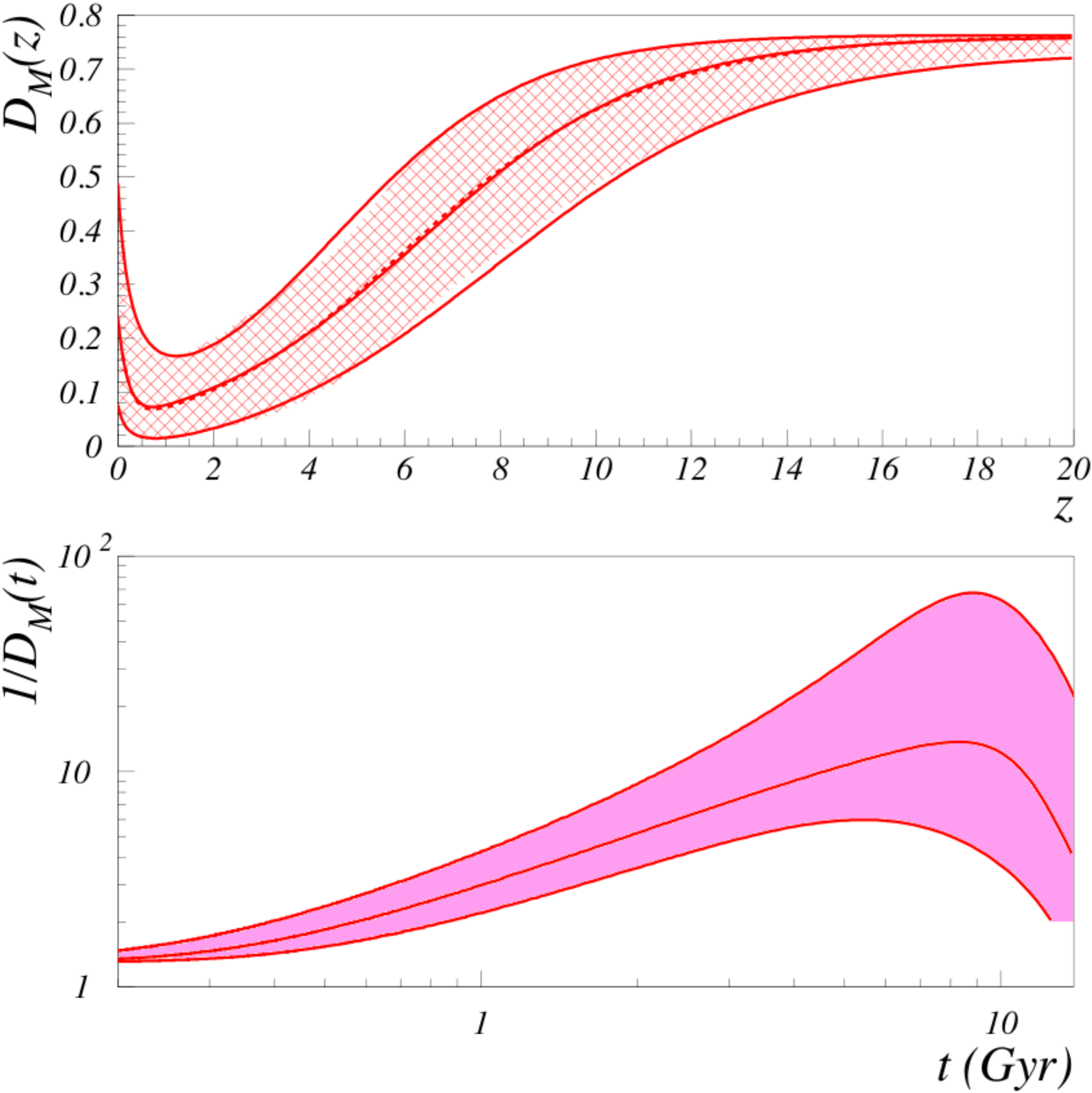}
\vspace{-0.4cm}
\caption{Top: depletion function $D_{M}$ vs redshift $z$, from Ref.~\cite{Deur:2017aas}.
%from the time-dependence of the fractions of baryonic matter contained in each type of large structure, {\it viz} elliptical or disk galaxies, groups, cluster and superclusters. 
The dashed line is the nominal $D_{M}(z)$ value, the hatched band being
its uncertainty. 
The three solid lines are parametrizations of the band envelope and nominal $D_{M}(z)$.
Bottom: inverse function $D^{-1}_{M}$ vs time $t$.}
\label{fig:invD(t)}
\vspace{-0.5cm}
\end{figure}

The SI-enhanced gravitational interaction should hasten the collapse of overdensities and the 
merging of primordial overdensities or of proto-galaxies, resulting in faster growth rates and a reduced need for dark matter.
The formalism  of Ref.~\cite{Deur:2017aas} can be used to quantitatively test this possibility.
%%%%%%%%%
Ref.~\cite{Deur:2017aas} points out that the SI-enhanced  binding inside massive systems is balanced by 
reduced gravitational field outside these systems, and shows how this reduction can explain 
without dark energy the large-$z$ supernova observations~\cite{large-z 98 data}.
To show this, the effective large-scale weakening of gravity was folded into a {\it depletion function} $D_M (z)$ that
quantifies the average suppression of gravity at large scales, with
$D_M=0$ indicating full suppression and $D_M=1$ none. In particular, 
the smallness of the density fluctuations and the homogeneity and isotropy of the early  universe
suppress SI effects at large $z$. Therefore   %$D_M(z)\scriptstyle{\underset{z \to \infty}{\longrightarrow}}$1. 
$D_M(z) \approx 1$ at large $z$.
In contrast, present times are characterized by a structured universe. If all fields were trapped in structures, 
%there would be no large-scale manifestation of gravity: 
then $D_M(z) \approx 0$ at small $z$. % $D_M(z)\scriptstyle{\underset{z \to 0}{\longrightarrow}}$0. 
However, SI effects tend to cancel in symmetric 
structures %~\cite{Deur:2009ya, Deur:2019kqi, Deur:2013baa, Deur:2017aas}
and the increasing ratio of elliptical over disk galaxies at late times makes $D_M(z)$ rise at small $z$.
$D_M (z)$ from Ref.~\cite{Deur:2017aas} is shown in the top panel 
of Fig.~\ref{fig:invD(t)}.
$D_M (z)$ appears in the derivation of the Friedmann %'s universe evolution 
equation as a factor to $G$, once the standard approximations of 
isotropy and homogeneity of the universe are lifted.
$D_M (z)$ is constructed using the time-dependent fractions of baryonic matter contained in each type of large structure
({\it viz} what fractions of the universe's baryonic matter are contained in galaxies, groups, clusters and superclusters)
and the time-dependent ratios of approximately spherical systems to non-spherical systems (e.g. the ratio of elliptical over disk and peculiar galaxies). %This later is needed because SI effects are suppressed in spherical systems.
Specifically, the amount of galaxies formed is modeled in~\cite{Deur:2017aas} with a Fermi-Dirac (FD) function of width equals to 
the characteristic period during which galaxies form. The limits of the function, 
0 and 1, express full SI effects and overall suppression of them, respectively. 
Since galaxies are but one of the types of large structures, the FD function is normalized by the 
relative amount of baryonic matter contained in galaxies. 
$D_M(z)$ is the product of the galactic FD function and that modeled for groups, 
clusters and superclusters. $D_M(z)$ also contains a correction term
that accounts for the suppression of SI effects in approximately spherically symmetric structures. 
This correction is proportional to the ratio of elliptical to other galaxies. Alternatively to FD functions,  
we also constructed $D_M(z)$ using linear functions, or convolutions of an Heavside function with 
a Gaussian function. Comparing these $D_M(z)$ to the one obtained with FD functions shows  
that dependence on the choice of function is small compared to the uncertainty band of $D_M(z)$.
%  
\iffalse
Equivalently for the present article, we could obtain  $D_M (z)$ from the large-$z$ supernova
data~\cite{large-z 98 data}, since the universe evolution accounting for $D_M (z)$ agrees well with these
data. This possibility emphasizes a connection between evidence for dark energy 
(here, supernova data) and evidence of dark matter (here, its role as seed for structure formations) that is hard 
to interpret in the dark universe paradigm ({\it viz} the $\Lambda$-CDM model), but is straightforward 
in the GR  SI approach. %\footnote{Another such connection is the cosmic coincidence~\cite{Zee}, dubbed so since it is not explained within $\Lambda$-CDM but is a trivial consequence of energy conservation in the GR's SI approach~\cite{Deur:2017aas}.}
\fi
%
%

Since $D_M $ quantifies the global large-scale suppression of gravity stemming from the strengthening 
of the structure internal binding, the strengthening
behaves like $D_M^{-1}$. 
That it is directly given by $D_M^{-1}$ is the simplest assumption, which we shall use. 
Since by definition $0 \leq D_M  \leq 1$, then $D^{-1}_M  \geq 1$, in agreement with the SI-increase of system binding. 
For example, the present value $D^{-1}_M  \approx 4.2$ (Fig.~\ref{fig:invD(t)}, bottom panel) 
implies that large structures are bound on average 4.2 times more than
a Newtonian analysis with only baryonic matter would indicate. 
%This is close to the ratio of DM/baryonic matter, about 27%/5%=5.4. That's probably not accidental but the details would need to be worked out.
%
Before calculating the structure growth accounting for  GR's, we
first recall the evolution equation of a baryonic overdensity $\delta(t)$ without SI.
It stems from the continuity and Euler equations, with Newton's potential used in the latter: 
\vspace{-0.1cm}
 \begin{equation}
\frac{d^2\delta(t)}{dt^2} +2\frac{dR(t)/dt}{R(t)}\frac{d\delta(t)}{dt} +(c_s(t)^2k^2-4\pi G \rho_0(t))\delta(t)  =0,
\label{eq:overdensity evol eq. 1}
\vspace{-0.2cm}
\end{equation} 
 where 
 %$t$ is the time, 
 $R$ is the overdensity radius, 
 $c_s$ the sound speed,
 $k$ the overdensity wavevector and
% $G$  the gravitational constant and
 $\rho_0$ the average density. 
 Since $D^{-1}_M$ represents the average increase of the gravitational binding inside systems, it factors the term 
 proportional to $G$, just like in Ref.~\cite{Deur:2017aas} $D_M$ factors $G$ in the Friedmann equation.
 Eq.~(\ref{eq:overdensity evol eq. 1}) thus becomes:
\vspace{-0.1cm}
 \begin{equation}
\frac{d^2\delta(t)}{dt^2} +2\frac{dR(t)/dt}{R(t)}\frac{d\delta(t)}{dt} +\big(c_s(t)^2k^2-4\pi GD^{-1}_M(t) \rho_0(t) \big)\delta(t)  =0.
\label{eq:overdensity evol eq. 2}
\vspace{-0.2cm}
\end{equation} 
%
%That $D^{-1}_M(z) \to 1$ at large-$z$ is compatible with the spherical symmetry of the Jeans collapse
%mechanism together with the suppression of SI for spherically symmetric systems.
 For the matter dominated epoch $c_s^2k^2 \ll 4\pi G \rho_0$ and for a flat universe, %(as observed), 
 Eq.~(\ref{eq:overdensity evol eq. 2}) becomes:
\vspace{-0.1cm}
 \begin{equation}
\frac{d^2\delta(t)}{dt^2} +\frac{4}{3t}\frac{d\delta(t)}{dt} - \frac{2D^{-1}_M(t)}{3t^2 }\delta(t)  =0.
\label{eq:overdensity evol eq. 3}
\vspace{-0.2cm}
\end{equation} 

\begin{figure}[!h]
\center
\includegraphics[width=0.40\textwidth]{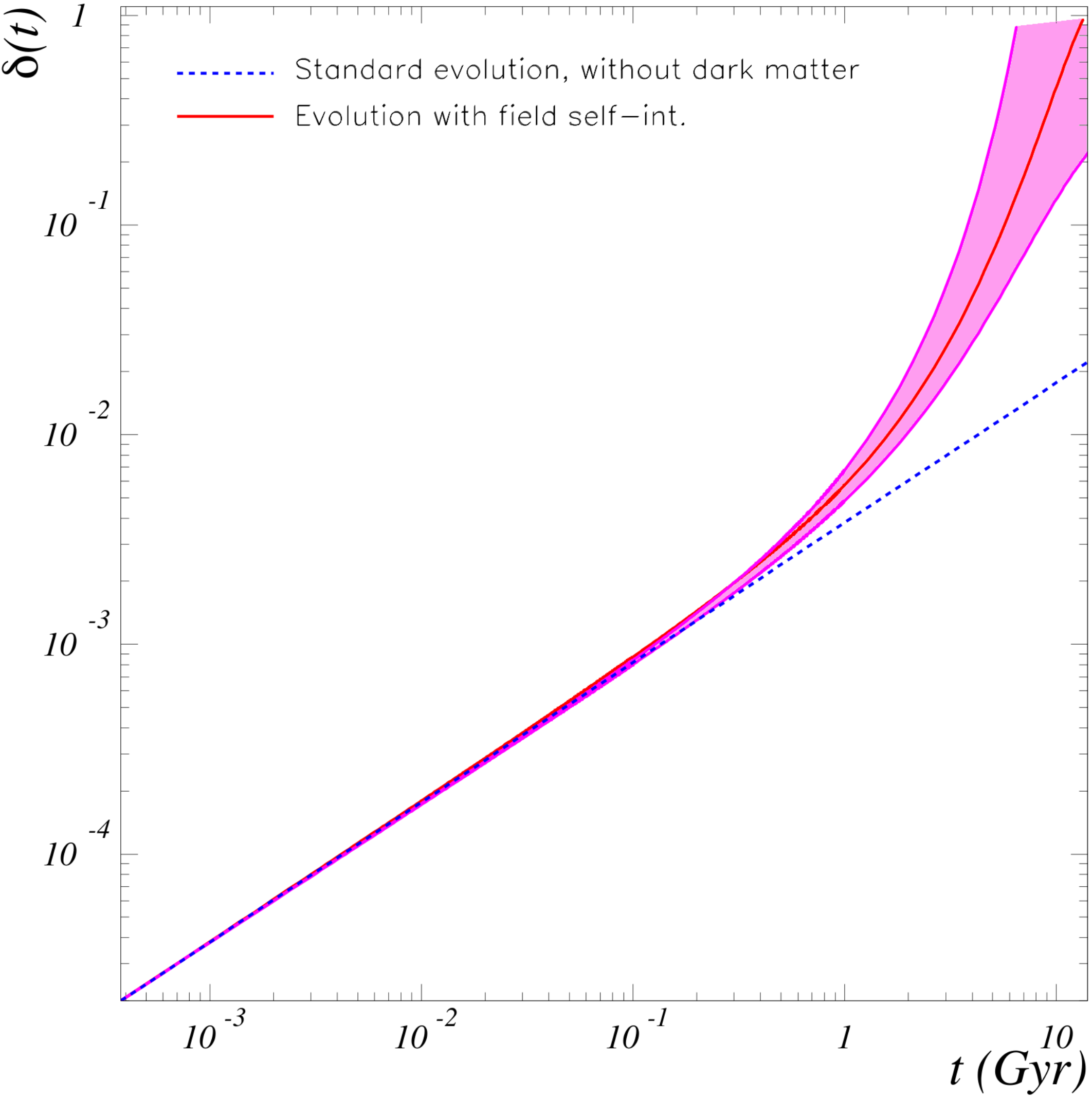}
\vspace{-0.4cm}
\caption{Time-evolution of a baryonic overdensity $\delta$. The initial value of $\delta$ at the recombination time, 
$t \approx 3.7$ Gyr, is $2 \times 10^{-5}$.  
The band shows the evolution including field SI effects, with the central solid line corresponding to the nominal 
$D_M^{-1}(t)$. The dashed line is the evolution without  SI. In neither case, dark matter has been assumed. 
}\label{fig:large struct. formation}
\end{figure}

The uncertainty band for $D_M(z)$ shown in Fig.~\ref{fig:invD(t)}
was obtained in Ref.~\cite{Deur:2017aas} with the usual procedure of propagating %the effects of 
the systematic uncertainties of the parameters on which $D_M(z)$ depends. Specifically, the widths and 
centers of the FD functions, the fractions 
of baryonic matter contained in the different structure types, and the 
ratio of elliptical to other galaxies, are systematically varied within ranges determined by
their respective uncertainties. This provides a distribution of $D_M(z)$ curves around the nominal $D_M(z)$.
The envelope of the distribution is taken as the uncertainty band on $D_M(z)$. The procedure conservatively 
assumes that the individual parameters' systematic uncertainties add linearly. 
With such procedure, the band 
cannot be expressed in closed form. %analytic expression. 
Since a closed form is convenient for the present work, we parameterized $D_M$ and found 
that the following expressions fit well the upper and lower limits of the band, respectively:
\vspace{-0.1cm}
\begin{eqnarray}
D_M (z)^{up}=0.84\bigg(0.9-\frac{1}{1.+e^{\sfrac{(z-4.3)}{2.1}}}+\frac{0.17}{z+0.3}\bigg);  ~~~
D_M (z)^{low}=0.80\bigg(0.915-\frac{1}{1.+e^{\sfrac{(z-7.9)}{2.9}}}+\frac{0.024}{z+0.2}\bigg).
\label{eq:Dep_band}
\vspace{-0.2cm}
\end{eqnarray}
The nominal $D_M (z)$ has a closed form %analytic expression 
but for convenience, we  parameterize it with a form similar to those above:
\vspace{-0.1cm}
\begin{equation}
D_M(z)^{nominal}=0.76\bigg(1-\frac{1}{1.+e^{\sfrac{(z-6.3)}{2.4}}}+0.25e^{-5z}\bigg).
\label{eq:Dep_nom}
\vspace{-0.2cm}
\end{equation} 
Eqs.~(\ref{eq:Dep_band}) and (\ref{eq:Dep_nom}) are traced by solid lines in Fig.~\ref{fig:invD(t)}, top panel. $D_M (z)$ is re-expressed with 
$t$ and inverted to obtain $D^{-1}_M(t)$, shown in bottom panel of Fig.~\ref{fig:invD(t)}. With it, 
we solved numerically Eq.~(\ref{eq:overdensity evol eq. 3}) using an initial overdensity  $\delta =2 \times 10^{-5}$ at 
$t \approx 0.37$ Myr ($z \approx 1100$). This corresponds to CMB temperature fluctuations of 55 $\mu$K, 
typical of the observed CMB data~\cite{CMB} which show temperature fluctuations reaching 75 $\mu$K and 
50 $\mu$K for the main and second acoustic  peaks, respectively.
%%%from Sparke & Galagher, page 346,  using T=2.73K for the average CMB temperature
The time evolution of $\delta(t)$ from $t =0.37$ Myr up to present time is shown in Fig.~\ref{fig:large struct. formation}. 
The solid line is the evolution including SI, Eq.~(\ref{eq:overdensity evol eq. 3}), and the band is the uncertainty due to $D_M^{-1}$.
The dashed line is the evolution without SI, Eq.~(\ref{eq:overdensity evol eq. 1}). %In both cases, no dark matter is assumed. 
The perturbation $\delta(t)$ is evolved as long as $\delta(t)<1$. Once $\delta(t) \sim 1$, Eq.~(\ref{eq:overdensity evol eq. 1}) 
breaks down and non-linear effects arise.
%see e.g. Weinberg Cosmology, or A. Loeb book "how did the first stars and galaxies formed, page 22.
%see also https://en.wikipedia.org/wiki/Structure_formation#Nonlinear_structure
%Note that  the evolution equations are perturbative, so it has to break down around $\delta(t) = 1$.
These non-linearities are of different origin than those induced by GR's SI since 
they also appear with Newton's gravity, a linear theory.
We do not need to consider here the non-linear regime that supersedes 
Eq.~(\ref{eq:overdensity evol eq. 1}) since the global perturbative treatment of Eq.~(\ref{eq:overdensity evol eq. 3}) 
is sufficient to conclude that due to SI increasing locally the 
action of gravity, overdensities can reach the observed present $\delta \gtrsim 1$ 
values, without need of dark matter. 

The picture for structure growth that emerges from the result showed in Fig.~\ref{fig:large struct. formation} is that 
at early times, SI did not influence the Jeans collapse mechanism since the initial overdensities were spherical. 
Mergings of overdensities would also not be 
significantly enhanced since density anisotropies would be too small to trigger the onset of SI. At later times, the %(baryonic) 
overdensities lose their spherical symmetry due to mergers and radiative energy dissipation. The SI-enhanced  internal binding
then accelerates local collapses. As anisotropies become denser, merging rates increase, especially for overdensities that
had so far retained their  spherical shape. 
The quantitative analysis shows that the SI-enhanced gravitational interaction is sufficient to form structures reaching the present-day densities, without requiring dark matter.  
We have considered here the time-evolution of $\delta$ and not the related subject of matter's spatial distribution. 
As shown in Ref.~\cite{Deur:2017aas}, the same formalism also
yields a position of the peak of the matter power spectrum of  $k_{eq}\simeq0.014$ Mpc$^{-1}$, 
%$k_{eq}=H_{0}\sqrt{2\Omega_{b}D_{M}(z=0)/a_{eq}}\simeq0.014$ Mpc$^{-1}$, 
%%%From Weinberg's Cosmology, Eq. 8.1.141 (page 412) and (1+Z_eq)=a0/a_eq=Omega_M/Omega_R (see p280), Z_eq~5400.
%%%Or also page 204 (or 301): Omega_R=/Omega_M a_eq/a0.
%%% H_0 \approx 73 km/s/Mpc, Omega_{b}=1, D_{M}(z=0) \approx 0.3; a_{eq} \apprrox 
%%%See Weinberg's Cosmology page 415 for the numerical estimate
in agreement with observations.
%Here, $H_0$ is the Hubble constant, 
%$\Omega_{b}$ the baryonic density fraction, and
%$a_{eq}$ the scale parameter at the matter-radiation equilibrium epoch ($z_{eq}\simeq3400$). 
%Again, no dark matter is assumed. 

\iffalse
% No dwarf galaxy problem ? We sketch it below but we should double check the reasoning.
This approach naturally solves the dwarf galaxy problem that arises in the $\Lambda-CDM$ paradigm~\cite{Klypin:1999uc}
since no dark matter is assumed and therefore there is no small dark matter halo formation to act as seed that would produce
too many dwarf galaxies. 
More specifically, the RMS overdensity fluctuation $\sigma_{of}$ at present time is calculated using $Omega_M$ 
without involving $D_{M}(z=0)$ since G does not enter the calculation of  $\sigma_{of}$. A $Omega_M \approx 0.045$, 
approximately 4 times smaller than if dark matter is included yields approximately  10\% less galaxies (Way not enough to solve the pb)
 %see pages 24-25...  A. Loeb book "how did the first stars and galaxies formed
 %see also Biney-Tremaine page 732
\fi

\section{Summary}

%Dark matter and dark energy consistently solve many cosmological observations that would otherwise be puzzling.
%This needed not be the case: as it happens often for a complex system consisting of interconnected components, 
%such as the universe, it may well be that various problems that challenge our understanding have independent causes but the
%interconnections misguide us in searching for a single origin of the problem. 
%Nevertheless, 
The consistency of the standard $\Lambda$-CDM model of the universe in explaining many observations that would be otherwise problematic is a 
compelling argument for the existence of dark matter and dark energy.
Yet, there are good reasons for studying alternatives to $\Lambda$-CDM, e.g. 
the lack of detection of dark particles, 
the dwindling support from theories beyond the standard model of particle physics, 
observations that challenge the dark matter model such as~\cite{McGaugh:2016leg},
lack of observations of $\Lambda$-CDM predictions such as the dwarf galaxy problem, or
the Hubble tension~\cite{Riess:2020sih}.
The credibility of an alternative approach is enhanced if, like for $\Lambda$-CDM, it can consistently explain the otherwise 
puzzling  cosmological observations.  
One alternative approach proposes that these observations are explained by the self-interaction 
of gravitational fields in General Relativity. It naturally explains the galactic rotation 
curves~\cite{Deur:2009ya}-\cite{Deur:2020wlg}, 
the supernovae observations suggestive of dark energy~\cite{Deur:2017aas},
the tight empirical relation between baryonic and observed accelerations~\cite{McGaugh:2016leg, Deur:2019kqi},
the dynamics of galaxy clusters~\cite{Deur:2009ya} and the Tully-Fisher relation~\cite{Tully:1977fu, Deur:2009ya, Deur:2017aas}.
The explanation is natural in the sense that a similar phenomenology is well-known in the context of QCD, a fundamental force
whose Lagrangian has the same structure as that of General Relativity. 
Crucially, no free  parameters are necessary, nor exotic 
matter or fields, nor modifications of the known laws of nature.
In this article, we checked whether the approach also explains the formation of large structures. 
We found that field self-interaction strengthens sufficiently the gravitational force so
that the small CMB inhomogeneities can grow to the density presently observed. Again, no free parameters were
needed: the function that globally quantifies the effect of field self-interaction had been previously  determined in Ref.~\cite{Deur:2017aas}
in the context of the {\it a priori} unrelated topic of dark energy. 

~\\
{ \bf Acknowledgments}
The author thanks D. Schwarz for initiating this work,
C. Sargent, S. \v{S}irca, B. Terzi\'c and X. Zhengh for useful discussions,
and the anonymous Phys. Lett. B reviewer for constructive comments that improved the manuscript.

\end{document}